\documentclass[twocolumn,showpacs,preprintnumbers,amsmath,amssymb]{revtex4}
\usepackage{amsmath}
\usepackage{graphicx}
\usepackage{amsfonts}
\usepackage{subfigure}
\usepackage{verbatim}
\usepackage{amssymb}%
\usepackage{verbatim}

\newcommand{\be}{\begin{equation}}
\newcommand{\ee}{\end{equation}}

\newcommand{\WDDensity}{$(4.8\times 10^{-3})$ pc$^{-3}$}
\newcommand{\WDDensityCite}{\cite{Hollberg:2008,Napiwotzki:2009}}
\newcommand{\WDBinaryFractionCite}{\cite{Lada:2006}}

\newcommand{\Msol}{M_{\odot}}

\usepackage{graphicx}
\usepackage{amssymb}
\usepackage{epstopdf}
\begin{document}
\title{Sources and technology for an atomic gravitational wave interferometric sensor}

\author{Michael Hohensee}
\email{hohensee@berkeley.edu}
\author{Shau-Yu Lan}
\author{Rachel Houtz}
\author{Cheong Chan}
\author{Brian Estey}
\author{Geena Kim}
\author{Pei-Chen Kuan}
\author{Holger M\"{u}ller}
\email{hm@berkeley.edu}
\affiliation{Department of Physics, 366 Le
Conte Hall, University of California, Berkeley, CA 94720-7300}

\begin{abstract}
We study the use of atom interferometers as detectors for gravitational waves in the mHz - Hz frequency band, which is complementary to planned optical interferometers, such as laser interferometer gravitational wave observatories (LIGOs) and the Laser Interferometer Space Antenna (LISA). We describe an optimized atomic gravitational wave interferometric sensor (AGIS), whose sensitivity is proportional to the baseline length to power of 5/2, as opposed to the linear scaling of a more conservative design. Technical challenges are briefly discussed, as is a table-top demonstrator AGIS that is presently under construction at Berkeley. We study a range of potential sources of gravitational waves visible to AGIS, including galactic and extra-galactic binaries.  Based on the predicted shot noise limited performance, AGIS should be capable of detecting type Ia supernovae precursors within 500 pc, up to 200 years beforehand.  An optimized detector may be capable of detecting waves from RX J0806.3+1527.
\end{abstract}
\maketitle

%\tableofcontents

\section{Introduction}
The production and propagation of gravitational waves is a central
prediction of the theory of General Relativity. Direct observation of such waves has been attempted using resonant bar detectors, which would be mechanically excited by passing gravitational waves, and using laser interferometers such as the laser interferometer gravitational wave observatories (LIGOs), in which the gravitational wave modulates the apparent distance between mirrors. These efforts have yet to detect gravitational waves, and improved interferometers such as LIGO-II are presently under construction, while the Laser Interferometer Space Antenna LISA interferometer is being developed.  These experiments hope to sense gravitational waves by focusing the search upon waves with lower frequencies (LISA), where many sources of gravitational waves have already been optically identified, and/or by increasing their overall sensitivity (LIGO-II).

\begin{figure}
\centering
\includegraphics[width=0.33\textwidth]{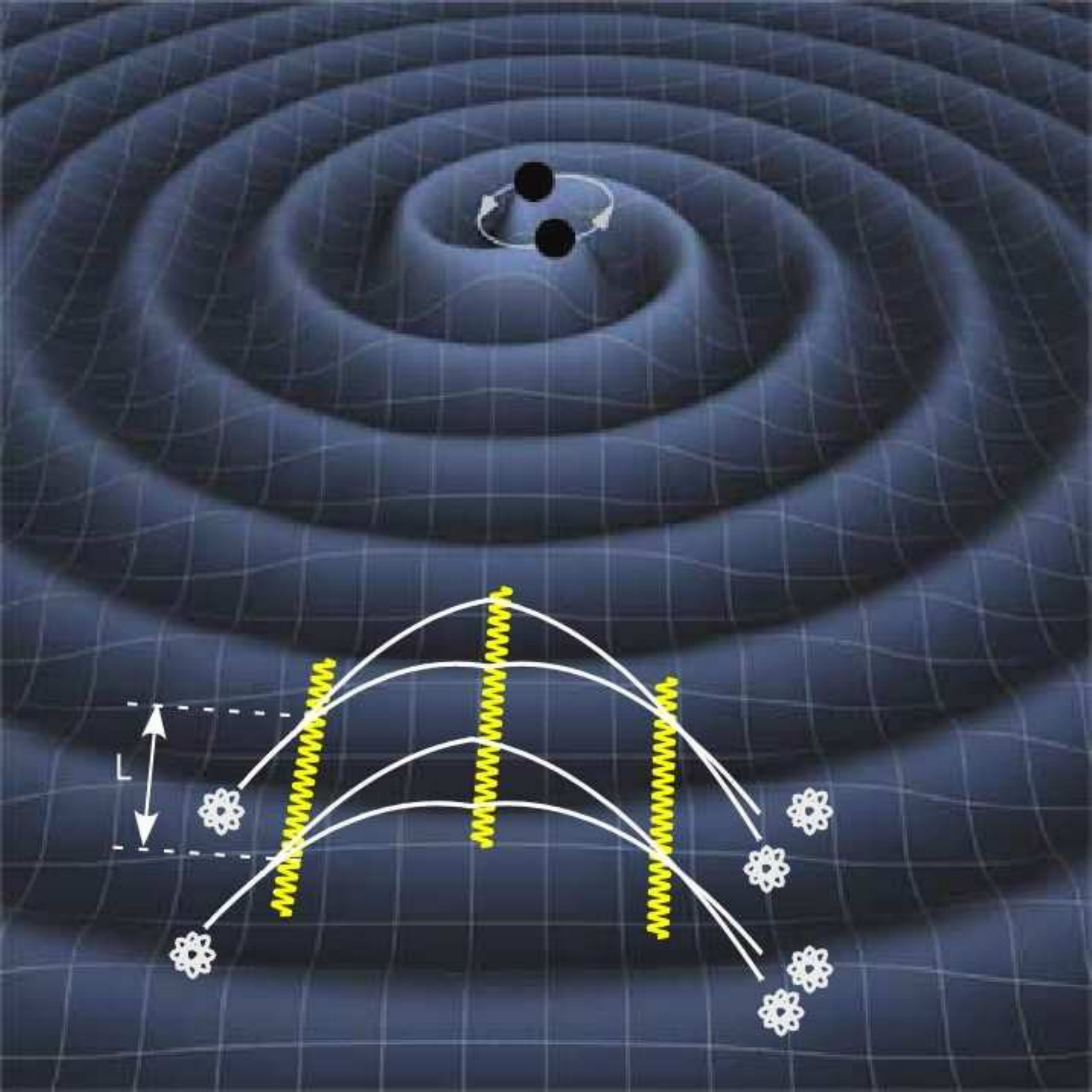}
\caption{\label{AGIS} AGIS: two atom interferometers located on
Earth or in space are separated by a distance $L$ and addressed by
the same laser radiation.}
\end{figure}

In this article, we study the application of atom interferometry to build an Atomic Gravitational Wave Interferometric Sensor (AGIS) \cite{GravWav}, see Fig.~\ref{AGIS}.  AGIS works on a similar principle to LIGO, replacing the measurement of the distance between macroscopic mirrors with measurement of the distance between freely falling atoms.  These atoms can approximate an inertial frame with high accuracy.  In contrast, the suspension of LIGO's macroscopic mirrors requires elaborate seismic isolation systems whose performance limits the sensitivity of LIGO at Hz-band or lower frequencies.

AGIS benefits from the fact that atoms are all alike, and have relatively few degrees of freedom, eliminating many sources of noise that are problems for LIGO.  There is no radiation pressure noise in AGIS, because each atom interacts with a fixed number of photons.  Thermal noise associated with the degrees of freedom of macroscopic mirrors and their suspension are likewise absent.  Radiation pressure and suspension thermal noise are the dominant low-frequency noise sources in Advanced LIGO.  These considerations suggest that AGIS might be more sensitive to gravitational waves at frequencies were the above noise sources are dominant, generally below 100 Hz, complementing LIGO and LISA. However, AGIS must contend with a higher level of shot noise, since the flux of atoms in AGIS is much lower than that of photons in LIGO.  This can nevertheless be mitigated by having each atom coherently interact with many photons.  Gravity gradient noise remains important to both AGIS and LIGO.

Thus, while AGIS seems interesting in the light of the above advantages, there are substantial challenges. The purpose of this study is to quantify some of these challenges, in particular the ones that concern the atomic physics aspects. We begin with a brief discussion of atom interferometers, and their use as key components of two AGIS detectors.  We present both a basic configuration, relying on reasonable extrapolations from technologies presently available, and an optimized AGIS where parameters are chosen to give maximal sensitivity for a given size of the apparatus.  This sensitivity scales favorably with the size of the apparatus.  We then turn to a brief survey of the potentially detectable sources of gravitational waves in and near the AGIS frequency band.   We give an overview of the state of the art of atom interferometry and on technologies that will need to be developed for AGIS; we also discuss some of the systematics that must be overcome. We hope this study will help to elucidate the promise and challenge of AGIS. Our study is mainly concerned with ground-based AGIS, though much of it will apply to space-based detectors \cite{Hogan:2010} just as well. 

\section{Atom Interferometers and AGIS\label{agissect}}

Light-pulse atom interferometry \cite{Pritchardreview} has been
used in measurements of local gravity \cite{Peters}, the
fine-structure constant \cite{Weiss,Wicht,Biraben}, gravity
gradients \cite{Snaden98}, Newton's gravitational constant
\cite{Fixler,Lamporesi}, a terrestrial test of
general relativity that is competitive with astrophysics
\cite{LVGrav,LVGravlong}, and a test of the gravitational
redshift with part-per-billion accuracy \cite{MuellerPetersChu}. The field has recently seen revolutionary advances in technology, improving the already impressive sensitivity of classical setups\cite{BraggInterferometry,Domen,SCI,BBB}. These technologies may be developed further into a tool for the detection of gravitational waves.

\begin{figure}[t!]
  \centering
  \includegraphics[width=0.45\textwidth]{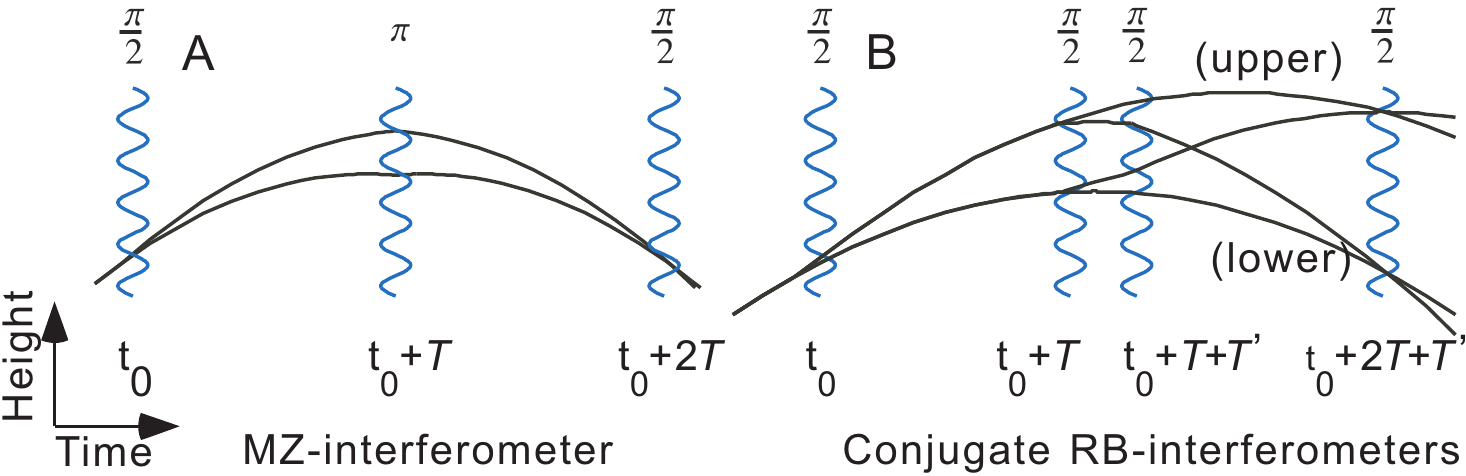}
  \vspace{-0.2cm}
  \caption{Left: Mach-Zehnder atom interferometer; Right: Ramsey-Bord\'{e}
Interferometer. Two conjugate interferometers are formed by
interfering either the upper or lower outputs of the second beam
splitter. Trajectories which do not interfere are not shown.}
  \label{MZRB}
\end{figure}

Figure \ref{MZRB} shows two basic atom interferometer configurations, the Mach-Zehnder and Ramsey-Bord\'e. In each case, an atomic matter wave is initially split by interaction with a
light pulse, which transfers the momentum of $n$ photons
with a probability near 1/2.  Subsequent pulses are used to redirect and
then recombine the trajectories. When the matter waves from both paths interfere, the
probability of observing the atom at a given output is given by
$(1+\cos \phi)/2$, where $\phi$ is the phase difference between matter waves in
both arms. This phase difference $\phi=\phi_F+\phi_I$ contains a
contribution due to the free evolution of the wave function,
$\phi_F=\Delta S_{\rm Cl}/\hbar$, given by the classical action $S_{\rm
Cl}$ as evaluated along the trajectories ($\Delta$ denotes the difference between both arms and $\hbar$ the reduced Planck constant). Another contribution
$\phi_I$ is because whenever a photon is absorbed (emitted), its
phase is added to (subtracted from) the one of the matter wave.

For a Mach-Zehnder interferometer, which will be the basic building block of AGIS, the leading order phase is given by
\begin{equation}
\phi_{\rm MZ}=n k g T^2,
\end{equation}
where $n$ is the number of photons transferred in the beam
splitters, $k$ is the laser wavenumber, $g$ the local
gravitational acceleration, and $T$ the pulse separation time
(Fig. \ref{MZRB}). For a Ramsey-Bord\'e interferometer, $\phi_{\rm
RB}=8n^2\omega_r T\pm n k g T^2$, where the plus and minus sign
refer to the upper and lower interferometer, respectively (Fig.
\ref{MZRB}) and $\omega_r\equiv \hbar k^2/(2M)$ is the recoil
frequency, where $M$ is the mass of the atom. The first term
arises because of the kinetic energy of the atoms due to the momentum transfer of the photons. This term is absent in Mach-Zehnder interferometers, in which both trajectories receive momentum from the photons at some time.  Recent technological advances in atom interferometry such as multiphoton Bragg diffraction, the use of simultaneous conjugate interferometers, and large momentum transfer using optical lattices may prove critical to the successful development of AGIS, and are reviewed below.

\subsection{AGIS}%\label{agissect}
\begin{table*}[ht!]
\centering \caption{\label{paramT} Parameters for AGIS that have
been assumed in the plots in the theory section of this paper.}
\begin{tabular}{cccc}
\hline\hline Parameter & Symbol & Basic & Optimized \\
\hline
Wavenumber & $k$ & $2\pi/852\,$nm & $2\pi/852\,$nm \\
Momentum transfer/$(\hbar k)$ & $n$ & 1,000 & 31,000 \\
Pulse separation time & $T$& 3\,s & 11\,s \\
Tube length & $L_{\rm Tube}$ & 1,000\,m & 3,000\,m \\
Separation & $L$ & $\approx L_{\rm Tube}$ & 1,200\,m \\
Atom throughput & $\eta$ & $10^{12}/$s & $3\times 10^{13}$/s \\
%Wavenumber & $k$ & $10^7/$m & $10^7/$m \\
Peak sensitivity & $h_{\rm rms}$ & $7\times 10^{-20}/\sqrt{\rm Hz}$ &  $1.3\times 10^{-22}/\sqrt{\rm Hz}$ \\
Low freq. sensitivity & $h_{\rm rms}^{\rm LF, opt}$ & $3\times 10^{-20}(\tfrac{{\rm Hz}}{\omega})^{2}\tfrac{1}{\sqrt{\rm Hz}}$ & $1.1 \times 10^{-23}(\tfrac{{\rm Hz}}{\omega})^{2}\tfrac{1}{\sqrt{\rm Hz}}$ \\
\hline\hline
\end{tabular}
\end{table*}

A basic AGIS setup previously discussed by Dimopoulos {\em et
al.} \cite{GravWav} is shown in Fig. \ref{AGIS}. Two atom
interferometers, separated by a distance $L$, are addressed by a common laser system. A passing gravitational wave with strain amplitude $h$ will modulate their distance $L$,
and thereby the differential phase of the atom interferometers.
The differential phase modulation will have an amplitude of \cite{GravWav}
\begin{equation}
\Delta\phi=2nkhL\sin^{2}(\omega T/2),
\end{equation}
where $\omega$ is the angular frequency of the gravitational wave.  The shot-noise limit for the sensitivity of AGIS is thus given by
\begin{equation}
h_{\rm rms} = \frac{1}{2nkL \sin^2(\omega T/2) \sqrt{\eta}},
\end{equation}
where $\eta$ is the average atom flux through the interferometer.
The wavenumber $k$ is determined by the atomic species used in the interferometer.  For Cesium, $k=2\pi/(852\text{ nm})$.  Detection of gravitational waves will require the atoms momentum to be coherently split by thousands of $\hbar k$.  Such large momentum splittings have yet to be experimentally demonstrated, but may be attainable by the use of accelerating optical lattices, as we discuss later in part~\ref{BBBSec}.

\begin{figure}[b] %  figure placement: here, top, bottom, or page
   \centering
   \includegraphics[width=0.47\textwidth]{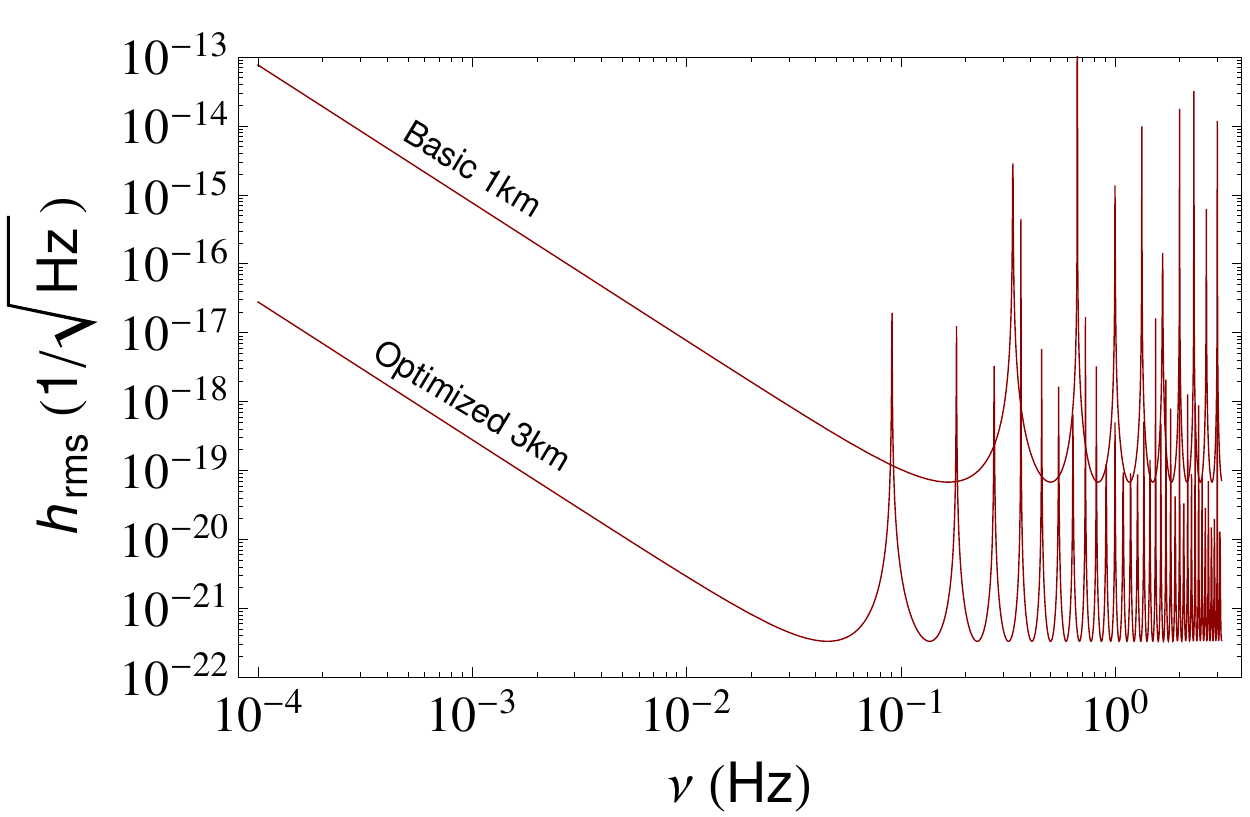}
   \caption{Root-mean-square strain noise per root Hertz for the two AGIS configurations described in the text and Tab.~\ref{paramT}.  \label{fig:strainnoise}}
\end{figure}

AGIS will also require high atom throughput $\eta$.  Atomic fountains using Raman sideband cooling have demonstrated launches of $2.5\times10^{8}$ state selected atoms at a three-dimensional temperature of $150$\,nK every two seconds~\cite{Treutlein}.  As discussed in part~\ref{sec:atomthroughput}, we estimate that this flux can be scaled to provide throughputs of $\eta = 10^{12}$\,atoms/s for AGIS, although shot noise limited detection of $10^{12}$ atoms has not yet been demonstrated.  We will use these parameters to define the characteristics of a ``basic'' AGIS detector, with a $L=1$\,km, as shown in Tab.~\ref{paramT}.  This yields a root mean square sensitivity $h_{\rm rms}=7\times 10^{-20}/\sqrt{\rm Hz}$ at odd multiples of the frequency $1/(2T)=1/6$\,Hz, illustrated in Fig.~\ref{fig:strainnoise}.  These parameters have been chosen because they appear feasible from extrapolation of the state of the art, at least in principle.  Additional sources of systematic error are discussed in more detail in part~\ref{sec:systematic}.

\subsection{Optimized AGIS}

We can also consider strategies for optimizing this shot noise limit in free-fall configured AGIS, based on physical limitations.

In the basic parameter set, we assumed an atom throughput of $10^{12}$/s. As explained in part~\ref{sec:atomthroughput}, this may be increased to $\sim3\times 10^{13}$/s
using 1\,kW of laser power for the two-dimensional magneto-optical trap (2D-MOT). This
higher flux can decrease shot noise by more than a factor of about 5.

In a free-fall configuration, the maximum pulse separation time $T$ is limited by the launch height of the atom
interferometer. Optimizing $T$ is crucial for reaching high sensitivity in the low-frequency limit $\omega\ll 2\pi/T$, where the sensitivity is given by
\begin{equation}
h_{\rm rms}^{\rm LF} = \frac{2}{nkL \omega^2 T^2 \sqrt{\eta}}.
\end{equation}
The length $L_{\rm Tube}$ of the vacuum tube for an AGIS with
parameters $L$ and $T$ is about $L_{\rm Tube} \approx L+ gT^2/2$, where $g$ is the acceleration of free fall.
Optimum sensitivity is reached when $T\rightarrow \sqrt{L_{\rm
Tube}/g}$, when the launch height is $L_{\rm Tube}/2$
\cite{GravWav}. The sensitivity is then
\begin{equation}
h_{\rm rms}^{\rm LF, opt}=\frac{4g}{nkL_{\rm Tube}^2 \omega^2 \sqrt{\eta}},
\end{equation}
or about 5 times better than with $T=3\,$s and the basic parameters in Tab. \ref{paramT}. As discussed in the outlook, it may be possible to further extend the pulse separation time by trapping the atoms between beam splitting pulses.

The size of the vacuum tube also limits the momentum transfer $n$, as the resulting spatial splitting $n v_r T$, where $v_r\sim 3.5\,$mm/s is the recoil velocity of cesium atoms at a wavelength of 852\,nm, must be accommodated. We conservatively assume that the full spatial splitting must be added to $L_{\rm Tube}$. Globally optimizing the low frequency sensitivity with respect to $T$ and $n$ yields
\begin{eqnarray}
T_{\rm opt}&=&\sqrt{\frac{2L_{\rm Tube}}{5g}}, \nonumber \\
n_{\rm opt}&=&\frac{2L_{\rm Tube}-gT^2}{4T v_r}, \nonumber \\
h_{\rm rms}^{\rm LF, opt}&=&\frac{25v_r\sqrt{5g} }{2kL_{\rm
Tube}^{5/2}\omega^2 \sqrt{2\eta}}.
\end{eqnarray}
This is 32 times better than the basic example of Tab.
\ref{paramT} for same tube length. Note the scaling of $h_{\rm rms}^{\rm LF,
opt}$ with $L_{\rm Tube}^{5/2}$.

The length $L_{\rm Tube}$ of the vacuum tube is limited mainly by
the height of available mine shafts, or other facilities that can
accommodate AGIS. 1\,km has been assumed for the basic scenario
in Tab. \ref{paramT}. In principle, the deep underground science and engineering laboratory (DUSEL) at Lead, SD, is deep enough to accommodate a 3\,km tube. Thus, $L_{\rm Tube}=3$\,km has been assumed for the optimized scenario, giving another increase in sensitivity by a factor of $3^{5/2}$.

All in all, the low-frequency shot noise limit of the ``optimized" AGIS (Tab. \ref{paramT}) is about 2,500 times as good as the basic one, a factor of 32 because of optimized $n,T$, a factor
of $3^{5/2}$ due to $L_{\rm Tube}$, times a final factor of
$\sqrt{30}$ due to the atom flux.

\section{Sources: Binary Inspirals}

At least a third of all stars are in binary or higher multiple
systems \cite{Lada:2006}.  Binary systems are believed to be
precursors to a variety of astrophysical phenomena of interest,
ranging from type Ia supernovae \cite{Ruiter:2009}, the formation
of neutron stars, and the evolution of supermassive black holes in
galactic nuclei \cite{Baker:2007}.  By detecting their emitted
gravitational waves, AGIS might contribute to our understanding of
such systems.  The complete gravitational wave spectrum emitted
over the full lifespan of any given binary system is generally difficult to
calculate, and particularly the final moments of the merger and ringdown
phase when the system deviates significantly from the Newtonian
approximation~\cite{Flanagan:1998}. In contrast, the long Newtonian inspiral which leads
up to the merger or collision of the two orbiting bodies is
well understood, and can be adequately modeled with simple analytic expressions
applicable to a range of compact objects, including white dwarfs, neutron stars, and black holes. A binary system loses energy in the form of gravitational
waves emitted with total luminosity \cite{Misner:1973}
\begin{equation}
L_{\rm
GW}=\frac{G^{4}}{c^{5}}\frac{32}{5}\frac{\mu^{2}M^{3}}{a^{5}}f(\varepsilon),
\end{equation}
where $\mu=M_{1}M_{2}/M$ is the system's reduced mass,
$M=M_{1}+M_{2}$ the total mass, $a$ is the distance separating the
two stars, and $G$ and $c$ are the familiar gravitational constant
and the speed of light.  $\varepsilon$ is the eccentricity of the
binary orbit, with $f(\varepsilon)$ given by \cite{Misner:1973}
\begin{equation}
f(\varepsilon)=\left[1+\frac{73}{24}\varepsilon^{2}+\frac{37}{96}\varepsilon^{4}\right]\left(1-\varepsilon^{2}\right)^{-7/2}.
\end{equation}
The angular frequency of the binary's orbit $\omega_b$ in the
Newtonian limit, valid for all inspirals considered here, is $\omega_b^{2}=GM/a^{3}$, allowing us to interchange $a$
and $\omega_b$ as needed.
Binaries in circular orbits ($\varepsilon=0$) emit only at the second harmonic, but for non-circular orbits, ($\varepsilon\neq0$), gravitational waves are also given off at higher harmonics $\omega_{{\rm GW},n}=n\omega_b$, for $n=2,3,4,\dots$.  The fraction $F_{n}(\varepsilon)$ of the total
luminosity $L_{\rm GW}$ emitted into the $n$th harmonic is given
by~\cite{Peters:1963}
\begin{equation}
F_{n}(\varepsilon)=g(n,\varepsilon)/f(\varepsilon),
\end{equation}
with
\begin{multline}
g(n,\varepsilon)=\frac{n^{4}}{32}\left\{\left[J_{n-2}(n\varepsilon)
-2\varepsilon J_{n-1}(n\varepsilon)+\tfrac{2}{n}J_{n}(n\varepsilon)\right.\right.\\ \left.\left. +2\varepsilon J_{n+1}(n\varepsilon)-J_{n+2}(n\varepsilon)\right]^{2}\right.\\
\left.+(1-\varepsilon^{2})\left[J_{n-2}(n\varepsilon)-2J_{n}(n\varepsilon)+J_{n+2}(n\varepsilon)\right]^{2}
\right.\\
\left.
+\tfrac{4}{3n^{2}}\left[J_{n}(n\varepsilon)\right]^{2}\right\}.
\end{multline}
As the binary loses energy via emission of gravitational waves,
the distance $a$ between stars in a circular orbit decreases
according to~\cite{Misner:1973}
\begin{equation}
a(t)=a_{0}(1-t/\tau_{0})^{1/4},\label{eq:adec}
\end{equation}
with
\begin{equation}
\tau_{0}=\frac{5}{256}\frac{a_{0}^{4}}{\mu M^{2}},
\end{equation}
and $a_{0}=a(t=0)$.  This implies that the orbital frequency and
gravitational wave luminosity of the system steadily increases in
the Newtonian limit, unless otherwise perturbed (\emph{i.e.} by
collision or coalescence of the two bodies).  Highly elliptical
orbits emit more strongly, and inspiral more rapidly.  Since most
of the extra energy is emitted at periastron, such orbits will
gradually circularize over time~\cite{Peters:1963,Pierro:1996}.

Gravitational waves can propagate with one of two orthogonal polarizations, ``$+$'' and ``$\times$''.  The power per solid angle emitted at a given polarization by a circular binary ($\varepsilon=0$) at an angle $\theta$ to the binary rotation axis is
\begin{align}
\frac{dP_{+}}{d\Omega}&=L_{\rm GW}\frac{5}{32} \frac{1}{2\pi}\left(1+\cos^{2}\theta\right)^{2}\\
\frac{dP_{\times}}{d\Omega}&=L_{\rm GW}\frac{5}{32} \frac{2}{\pi}\cos^{2}\theta.
\end{align}
The ability of a detector to see a source emitting one or both
polarizations depends upon the orientation of the source relative
to the distance vector to the detector, as well as the detector's geometry.  For the purposes of estimating which objects AGIS might detect, we will assume that the detector is optimally oriented to detect the $+$-polarized mode from any given source, and that the emitting binaries' rotation axes are rotated by $\pi/3$ from the distance vector linking them to the detector.  The first assumption maximizes the signal in the detector, since $\frac{dP_{+}}{d\Omega}\geq\frac{dP_{\times}}{d\Omega}$ for all $\theta$, while the second simplifies our analysis, since the power emitted into the $+$ mode is then equal to that which would be supplied by an isotropic and unpolarized source.

In the low energy density limit, where their energy density is insufficient to cause significant self-gravitation effects, gravitational waves propagate as solutions to the conventional wave equation.  For a monochromatic wave with intensity $\Phi$ and frequency $\omega$, the dimensionless amplitude $h$ of the wave is given by
\begin{equation}
\Phi=\frac{c^{3}}{G}(2\pi\nu)^{2}h^{2}.
\end{equation}
The gravitational wave is a disturbance of the metric of spacetime itself, acting tidally to alternately stretch and compress the distribution of matter and energy transversely to its direction of propagation~\cite{Baker:2007}.  For freely falling objects separated by a distance $L$, passage of an appropriately polarized gravitational wave with amplitude $h$ causes the separation to vary sinusoidally with an amplitude of $\delta L=hL/2$.  Interferometric detectors such as AGIS make precise measurements of such variations, and are thus sensitive to the amplitude of the gravitational wave, rather than its intensity.  For the $+$-polarized mode emitted at an angle $\pi/3$ from the quadrupole axis of the source, the amplitude $h$ at a distance $r$ can be related to its total gravitational wave luminosity by
\begin{equation}
\Phi(r)=\frac{c^{3}}{G}(2\pi\nu)^{2}h^{2}=\frac{1}{2}\frac{L_{\rm GW}}{4\pi r^{2}},
\end{equation}
so that the metric strain amplitude $h$ is given by
\begin{equation}\label{eq:rhmag}
h=\frac{1}{2\pi\nu r}\sqrt{\frac{G L_{\rm GW}}{8\pi c^{3}}}
\end{equation}
Note that since our detectors are sensitive to the strain
amplitude $h$, rather than the energy density $\propto h^{2}$, the detected signals drop as
$1/r$.  A useful distance-independent definition of the ``magnitude''
of a particular harmonic of a gravitational wave source is thus most conveniently
defined as $r h$~\cite{Kopparapu:2007}.

To determine whether the magnitude of a given source of gravitational waves is sufficient for AGIS to detect, we convert our detector's $h_{\rm rms}/\sqrt{\text{Hz}}$ noise to a noise equivalent magnitude at the distance of the source.  This noise equivalent magnitude at a distance $r$ after an integration time $T$ is given by
\begin{equation}
(rh)_{\rm nem}(\nu)=\frac{r}{\sqrt{T}}h_{\rm rms}(\nu).
\end{equation}
Note that detection may not be reliably accomplished until a significant amount of additional time has elapsed, depending upon what signal to noise ratio (SNR) threshold is required.  For independent detection of gravitational waves against a Gaussian noise background, SNRs of 5 are typically considered necessary, while for significantly non-Gaussian backgrounds, the SNR threshold may be higher.  For joint-detection schemes, where some parameters of the emitted waves can be determined by {\emph e.g.} optical observation, a somewhat lower SNR may be acceptable in some cases.  Little is known about the noise backgrounds applicable to AGIS, so in what follows, we will simply plot the integrated noise equivalent magnitude, and, assuming a Gaussian noise spectrum, require a SNR of 5 for detection. Objects within a distance $r$ from the detector whose gravitational wave emissions lie above the plotted noise equivalent magnitude spectrum on the $rh$ vs. $\nu$ plot will be detectable with an SNR of 5 after the observation period $T$.

Detection of a gravitational wave requires continuous observations
over at least one, and typically many, wave periods.  The time
required to accumulate a sufficient number of cycles to
distinguish the gravitational wave signal from the detector's
background noise may in some cases exceed the lifetime of the
source, precluding detection.  To determine whether this is the
case, we use Eq. \eqref{eq:rhmag} and Eq. \eqref{eq:adec}, to relate the
magnitude $rh$ of a given $\varepsilon=0$ binary to the time
$\tau$ required to double its frequency.  This relation is
\begin{equation}
(rh)_{\rm double}=\frac{\sqrt{5}\left(8-2^{1/3}\right)c}{4096\pi^{7/2}\tau\nu^{2}},\label{eq:rhdouble}
\end{equation}
where $\nu=2\pi\omega=2 \omega_b/(2\pi)$ is the initial frequency of the emitted
(second harmonic) waves.  Note that the form of Eq. \eqref{eq:adec} is such that the time required for a binary to double its emission frequency is far greater than the lifetime remaining to it after its frequency has doubled.  This characteristic doubling time is
independent of the constituents of the system damped by
gravitational wave emission, and may therefore be broadly applied
to the evolution of compact binaries composed of white dwarfs, neutron
stars, or black holes.
Using Eq. \eqref{eq:rhdouble}, we can easily determine whether a given source of gravitational waves will survive longer than the time required to observe it.

\subsection{Compact Galactic Binaries}
Compact inspiraling binary systems located within our galaxy
constitute a promising source of gravitational waves which may be
detectable by AGIS.  Such systems may involve white dwarfs (WD),
neutron stars (NS), as well as black holes (BHs). We note that the discussion below assumes that AGIS will be shot-noise limited at frequencies near 0.01\,Hz. This poses the challenge of overcoming gravity gradient noise, as will be discussed later.

\subsubsection{White Dwarf Binaries}

The number density of galactic white dwarf stars is estimated to
be \WDDensity~\WDDensityCite, of which at least a third may be
expected to be involved in binary or some higher multiplicity
systems~\WDBinaryFractionCite.  Although WD binaries are
comparatively weak emitters of gravitational waves, their
comparative ubiquity and proximity to Earth makes them easily
detectable by space-based detectors like LISA~\cite{Baker:2007},
and also potentially observable by ground-based AGIS detectors.  Other ground-based detectors such as LIGO are primarily sensitive to signals above $\sim 10$ Hz.  WD-WD and WD-NS binaries are insufficiently compact to exist at such frequencies, and are thus not detectable by LIGO.
\begin{figure*}[t] %  figure placement: here, top, bottom, or page
   \centering
   \includegraphics[width=\textwidth]{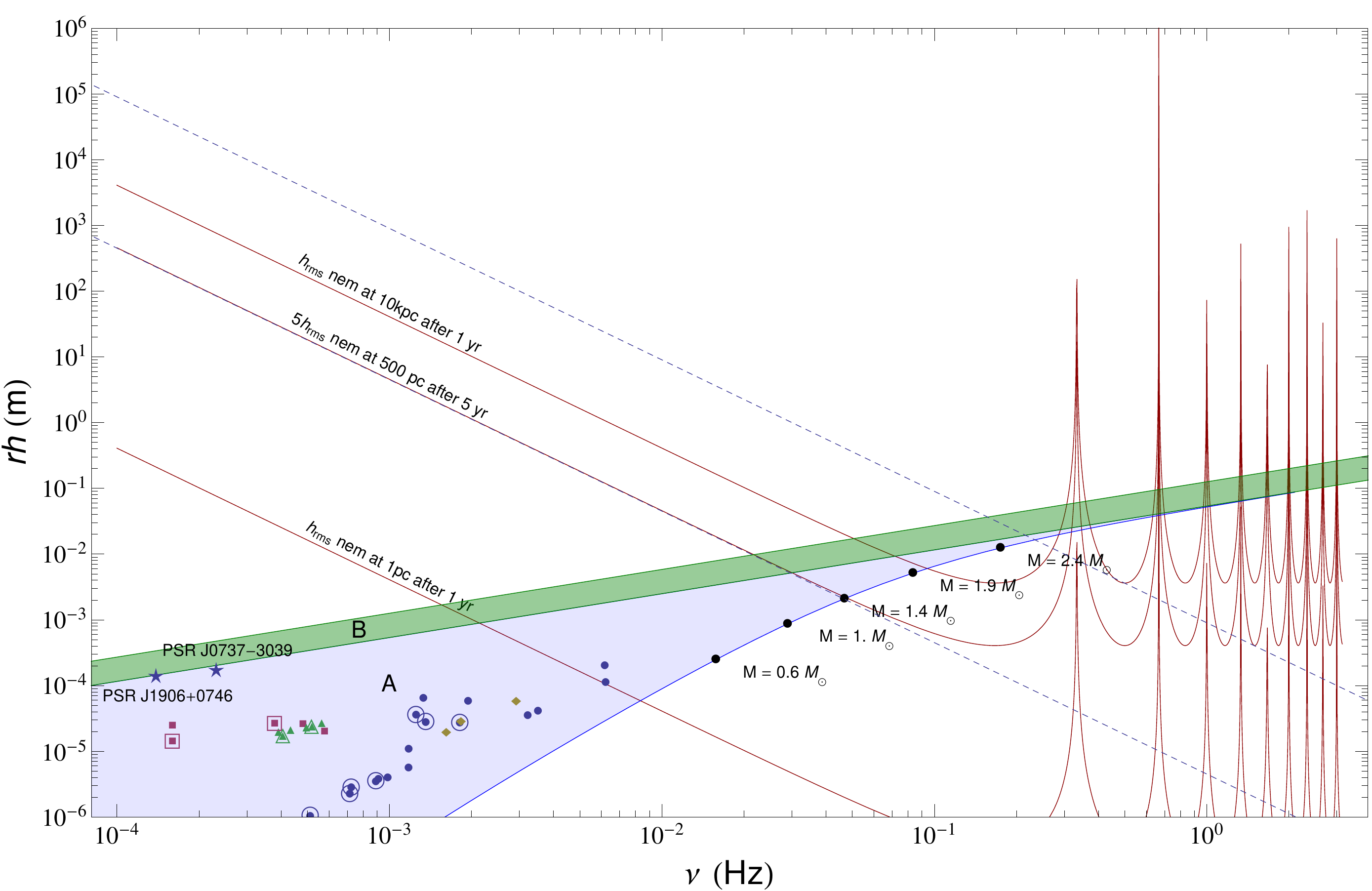}
   \caption{White dwarf and neutron star binaries. Gravitational strain magnitudes of potential WD-WD binaries (blue-shaded region A)~\cite{Kopparapu:2007} and NS-NS binaries
   (green-shaded region B).  Downward sloping solid lines indicate the AGIS noise equivalent strain magnitude spectrum at the indicated distances per root observation time for the basic detector parameters in Tab.~\ref{paramT}.  Objects with gravitational strain magnitudes larger than the equivalent noise magnitude $h_{\rm rms}$ at a particular distance will lead to a signal to noise ratio (SNR) of 1 after the indicated period of integration. Sources which produce a SNR of 5 lie above the $5h_{\rm rms}$ curves and are potentially detectable by AGIS.     Black
   circles on the boundary of the WD-WD binary region indicate points at
   which the components of WD-WD systems with total mass $M$ come into
   contact with one another.
   Downward-sloping dashed lines represent the $rh$ threshold above which inspiraling sources must
   double their emission frequency in less than 200 years (lower line) or one year (upper line).  Points
   in the lightly shaded region represent the gravitational strain
   magnitudes of the binary systems which make up the LISA verification
   sources, as given in~\cite{Stoeer:2006} and~\cite{NelemansWiki}, and also the more recently studied HM Cancri system~\cite{Roelofs:2010}.  These systems
   which include AM CVn systems (circles), detached WD binaries (squares),
   ultra-compact x-ray binaries (diamonds), and cataclysmic variables (triangles).
   Sources marked with circles are believed to be within 500 pc of the Sun.}
   \label{fig:WDPlot}
\end{figure*}

In contrast, an analysis~\cite{Kopparapu:2007,Kopparapu:2009} of the gravitational
wave strain magnitude $r h$ of WD-WD and WD-NS binaries, shown in Fig.~\ref{fig:WDPlot}, reveals that AGIS detectors
are sensitive to WD-WD binaries with period 200 s or below at a
distance of 1 pc, and systems with period 27 s and below at a
distance of 10 kpc.  This is a range sufficient to detect most
such fast binaries residing the nearer half of the galaxy.  The blue-shaded region A in Fig.~\ref{fig:WDPlot} shows the range of magnitudes and frequencies at which WD-WD and WD-NS binaries can emit, taking the Chandrasekhar mass to be $M_{ch}=1.44M_{\odot}$, while the green-shaded region B indicates the magnitude of NS-NS binaries, where the neutron star masses are assumed to be between $M_{ch}$ and $2.4M_{\odot}$.

A recent study~\cite{Ruiter:2009} suggests that type Ia supernovae
arising from WD-WD binaries with total mass greater than
$1.4M_{\odot}$, where $M_{\odot}$ is the mass of the sun, occur in our galaxy at a rate of $1\times 10^{-3}$
yr$^{-1}$.  Figure~\ref{fig:WDPlot} shows that basic AGIS (see Tab. \ref{paramT} and below) should be able to detect gravitational radiation from Ia supernova precursors with SNR greater than 5 at ranges exceeding 500 pc after five years of integration, and is sensitive to them as early as 200 years prior to the pending collision.

AGIS may also be sensitive to gravitational waves at higher
harmonics that are emitted from systems with even longer periods,
provided they have sufficiently eccentric orbits.  Although most
galactic WD binaries are expected to be in nearly circular orbits,
recent simulations~\cite{Willems:2007} suggest that binaries
residing in nearby globular clusters may have higher
eccentricities, due to interactions with nearby stars.  There are
at least 79 such globular clusters within 10 kpc of the
Sun~\cite{Willems:2007,Harris:1996}.  Further study will be
required to determine whether such globular clusters are likely to
contain binaries with significant eccentricities in the AGIS
detection band.

\begin{figure} %  figure placement: here, top, bottom, or page
   \centering
   \includegraphics[width=0.45\textwidth]{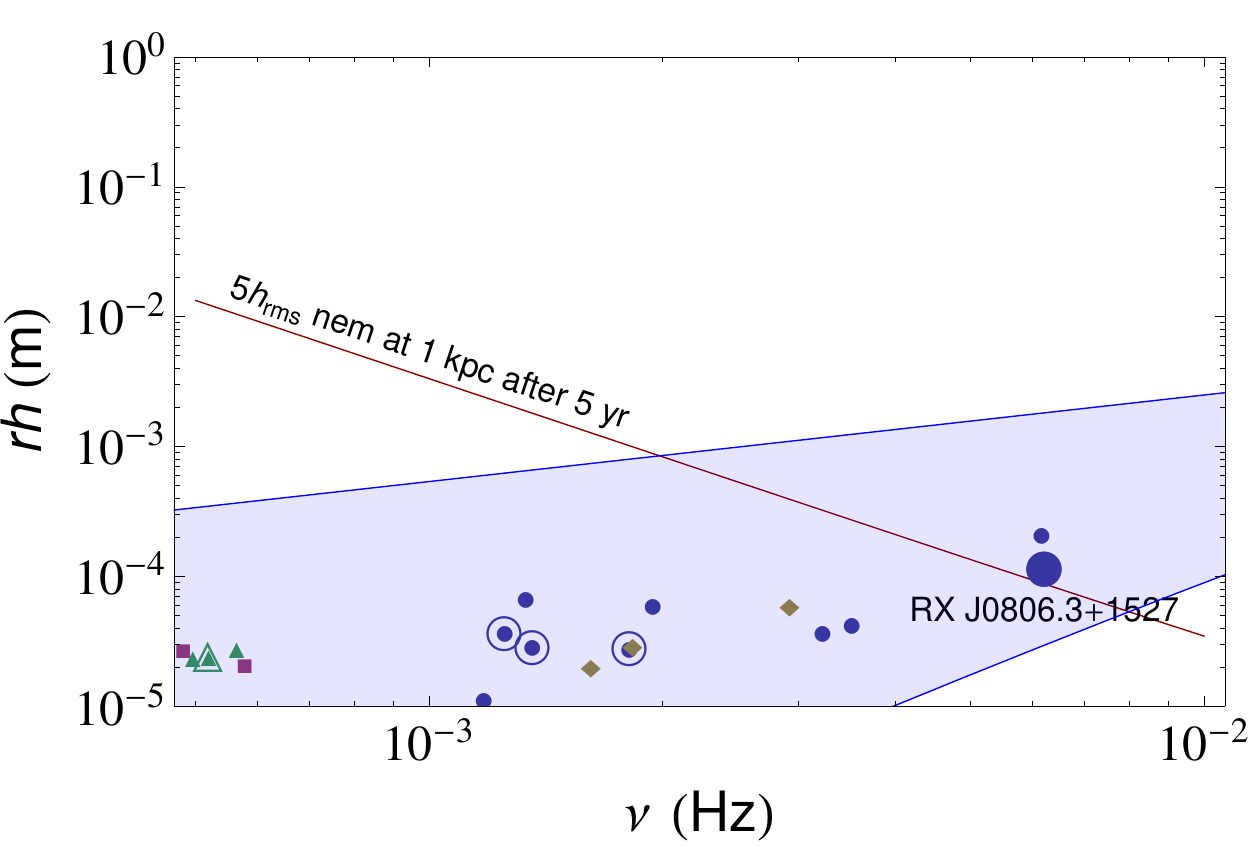}
   \caption{Enlargement from Fig. \ref{fig:WDPlot}. $5$ times the noise equivalent strain magnitude for an optimized $L=3$ km AGIS detector at 1 kpc.
   The enlarged point indicates a known source, RX J0806.3+1527, which would produce a signal with SNR greater than 5 after one
   year of integration.}
   \label{fig:ExtremeSensitivity}
\end{figure}

Although the reference AGIS detector discussed above is not
sufficiently sensitive to detect any presently known compact
binaries, an optimized detector with the optimized parameters
listed on Table \ref{paramT} would be.
Figure~\ref{fig:ExtremeSensitivity} depicts five times the noise equivalent
strain magnitude of the shot noise in such a detector at 1 kpc,
relative to the magnitude of the ``brightest'' known binaries.
This upgraded AGIS detector could detect gravitational waves from
RX J0806.3+1527 with SNR greater than 5 in five years~\cite{NelemansWiki}.

\subsubsection{Neutron Star and Low to Intermediate Mass Black Hole Binaries}

\begin{figure*}%  figure placement: here, top, bottom, or page
   \centering
   \includegraphics[width=\textwidth]{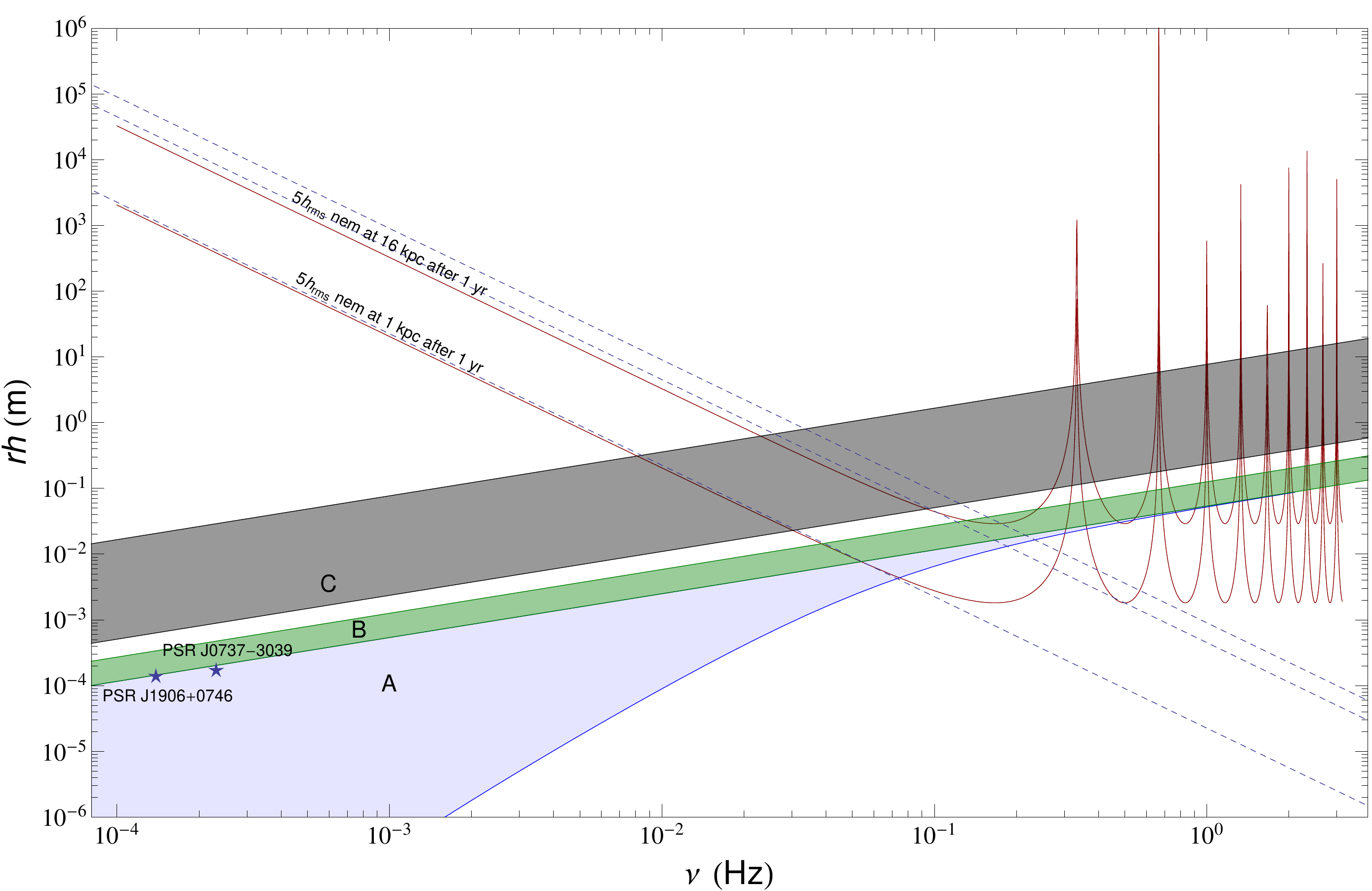}
   \caption{Neutron star binary sources.  Detectability of NS-NS binaries (with masses
   between $1.4\Msol$ and $2.6\Msol$) and those involving black holes
   ($M\leq 10\Msol$).  The upper shaded region C encompasses BH-BH and BH-NS
   binaries composed of a $10 \Msol$ BH with a partner whose mass lies between
   $1.44\Msol$ and $100\Msol$.  The three downward sloping dashed lines indicate
   the thresholds above which a source's orbital frequency doubles in less than 100 years
   (lower), two years (middle) and one year (upper). As PSR J0737-3039 lies approximately 600 pc away from Earth, GWs it emits will, in
   about 84 million years, exceed the AGIS's background noise after one year of observation,
   some 500 years before entering LIGO's detection band.}
   \label{fig:NSBHPlot}
\end{figure*}

NS-NS, NS-BH, and BH-BH binaries are attractive sources of
gravitational waves that AGIS might study.  These systems are
ultra-compact, and can in many cases exhibit a simple slow
inspiral over AGIS's entire detection bandwidth.  They are also
massive, making them significantly easier to detect at a given
distance as compared to systems involving WD stars.  Unfortunately,
they are also much less common.  There are at present at least five known
galactic NS-NS binaries that are expected to merge within several
hundred Myr: PSR J1756-2251, PSR J1906+0746, PSR B1534+12, PSR B1913-16, and PSR J0737-3039~\cite{Faulkner:2004,Lorimer:2006,Burgay:2003,Kalogera:2004}.  Of these five, only PSR J1906+0746 and
PSR J0737-3039 have sufficiently short periods to come within
three decades of AGIS's peak sensitivity bandwidth.  As
illustrated in Fig.~\ref{fig:NSBHPlot}, the reference AGIS
configuration considered here may be capable of detecting all
NS-NS binaries with SNR 5 within 1 kpc that have less than 40 years before coalescence after one year's observations.  The expected coalescence rate for our galaxy is not well known; one estimate is at about $10^{-5}$\,yr$^{-1}$ \cite{Freitas}.

The plot in Fig.~\ref{fig:NSBHPlot} of $5$ times the noise equivalent magnitude at 16 kpc after a year's integration of the basic AGIS considered here shows that one year's observations may be sufficient to detect most galactic NS-NS binaries with total mass greater than $4\Msol$ with SNR $>5$ at least a year before merger.  Since $16 \text{ kpc}/\sqrt{\text{month}}\simeq 50 \text{ kpc}/\sqrt{\text{yr}}$, the same curve indicates that binaries
composed of a $10\Msol$ black hole and a neutron star, or more massive partner, may be detectable more
than two years from merger anywhere in the galaxy or Large Magellanic Cloud after one year
of observations.

Recently, an intermediate-mass black hole (IMBH) between 500 and $10^4$ solar masses has been observed in the galaxy ESO 243-49 \cite{Farrell}, about 92 Mpc from Earth.  Inspirals of $10\Msol$ BH around this IMBH around this BH should be detectable by the optimized AGIS a month or more so before merger, as indicated in region H of Fig.~\ref{fig:EMCPlotExtreme}.  If the mass of the IMBH exceeds $2000\Msol$, NS and WD inspirals may also be detectable.  It must be noted, however, that this particular BH is believed to be the stripped nucleus of a small dwarf galaxy, and may no longer have a significant number of compact partners.  Its existence nevertheless suggests that there may be more IMBHs in the vicinity, making them promising candidate sources.

\subsection{Inspirals at the Galactic Core}
\begin{figure*} %  figure placement: here, top, bottom, or page
   \centering
   \includegraphics[width=\textwidth]{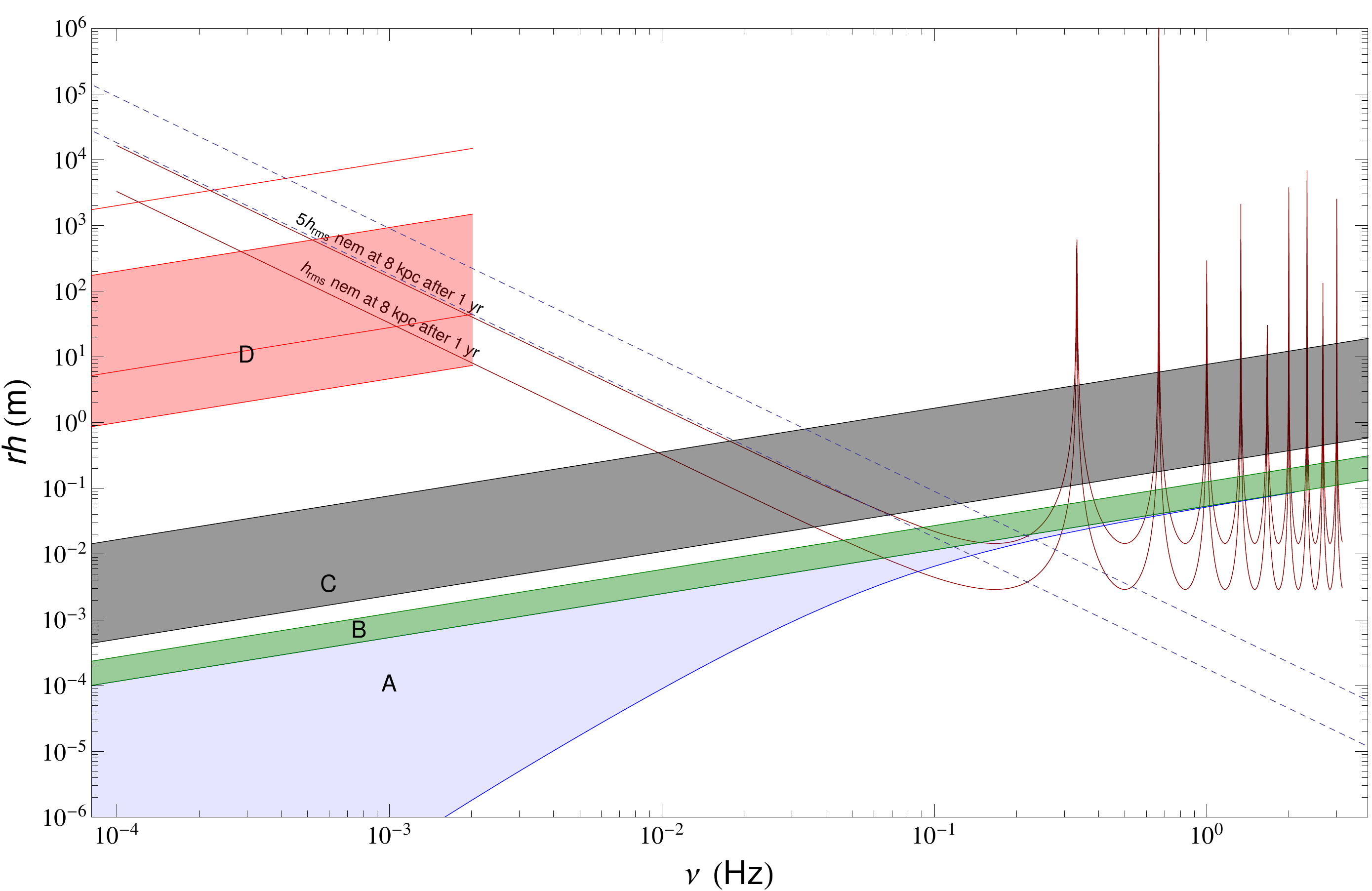}
   \caption{Galactic core sources. The red-shaded region D at the center left represents
   the gravitational strain magnitudes of extreme mass capture
   inspirals which may occur in the galactic core.  This involves
   a black hole with $M_{{\rm BH}}=4\times 10^{6}M_{\odot}$ and
   in-falling compact objects with masses ranging from $0.5M_{\odot}$
   (which form the region's lower bound) to $100 M_{\odot}$ (forming the region's upper bound).  The line which parallels the shaded region's upper boundary indicates the signal generated by a $10^{3}\Msol$ black hole in orbit about the larger black hole, while the line inside the shaded region indicates the signal generated by a similarly orbiting $3\Msol$ neutron star.   As before, the two dashed blue lines indicate the point at which inspiraling bodies double their frequency: in less than 5 years (lower dashed line) or in less than one year (upper dashed line).}
   \label{fig:CorePlot}
\end{figure*}

At the center of the galactic core, approximately $8$ kpc distant
from the Sun, there is believed to be a supermassive black hole
with total mass $M=4\times 10^{6}\Msol$~\cite{Ghez:2005}.  A basic shot noise limited AGIS detector would be capable of detecting inspiral waves from compact objects in close orbit about the core, provided that such objects exist and are close to merger.  As indicated in
Fig.~\ref{fig:CorePlot}, AGIS is sensitive with SNR $=5$ to waves emitted from 
$M=20\Msol$ black holes more than 3 years before coalescence with
the galactic core, and to the waves emitted by a $M=3\Msol$
neutron star in its final year with slightly reduced SNR. Circular inspirals involving the galactic core will not emit at frequencies
exceeding $\nu_{\rm max}$, given by
\begin{equation}
\nu_{\rm max}=\frac{c^{3}/\pi}{8GM_{\rm total}}\simeq 2 \text{
mHz},
\end{equation}
for $M_{\rm total}\simeq 4\times 10^{6}\Msol$.  Inspiral waves at higher frequencies are precluded as they would require the orbiting bodies to be within 2 gravitational radii of one another, where the Newtonian approximation is expected to be invalid. Inspirals from
systems with lower total mass can form closer orbits, and thus
emit at higher frequencies. Such low mass black holes are also
expected to abound in the vicinity of the galactic core, with
populations of the central parsec estimated to be of order
$10^{4}$~\cite{Freitag:2006,Miralda-Escude:2000,Morris:1993}.  Any such binaries which will merge in $\sim 5$ years may be
detected at $8$ kpc. Estimates for the capture rate for a given black hole span a disquietingly large range from $5\times 10^{-9}$\,yr$^{-1}$ \cite{Hopman} and $10^{-5}$\,yr$^{-1}$ (and, for anomalously heavy neutron stars even $10^{-4}$\,yr$^{-1}$) \cite{Freitag:2001} due, in part, to the lack of realistic agreed-on models for the structure of galactic nuclei. This puts the detection probability for AGIS anywhere between order-unity and $10^{-4}$ per year.

\subsection{Extragalactic Extreme Mass Ratio Inspirals}
\begin{figure*} %  figure placement: here, top, bottom, or page
   \centering
   \includegraphics[width=\textwidth]{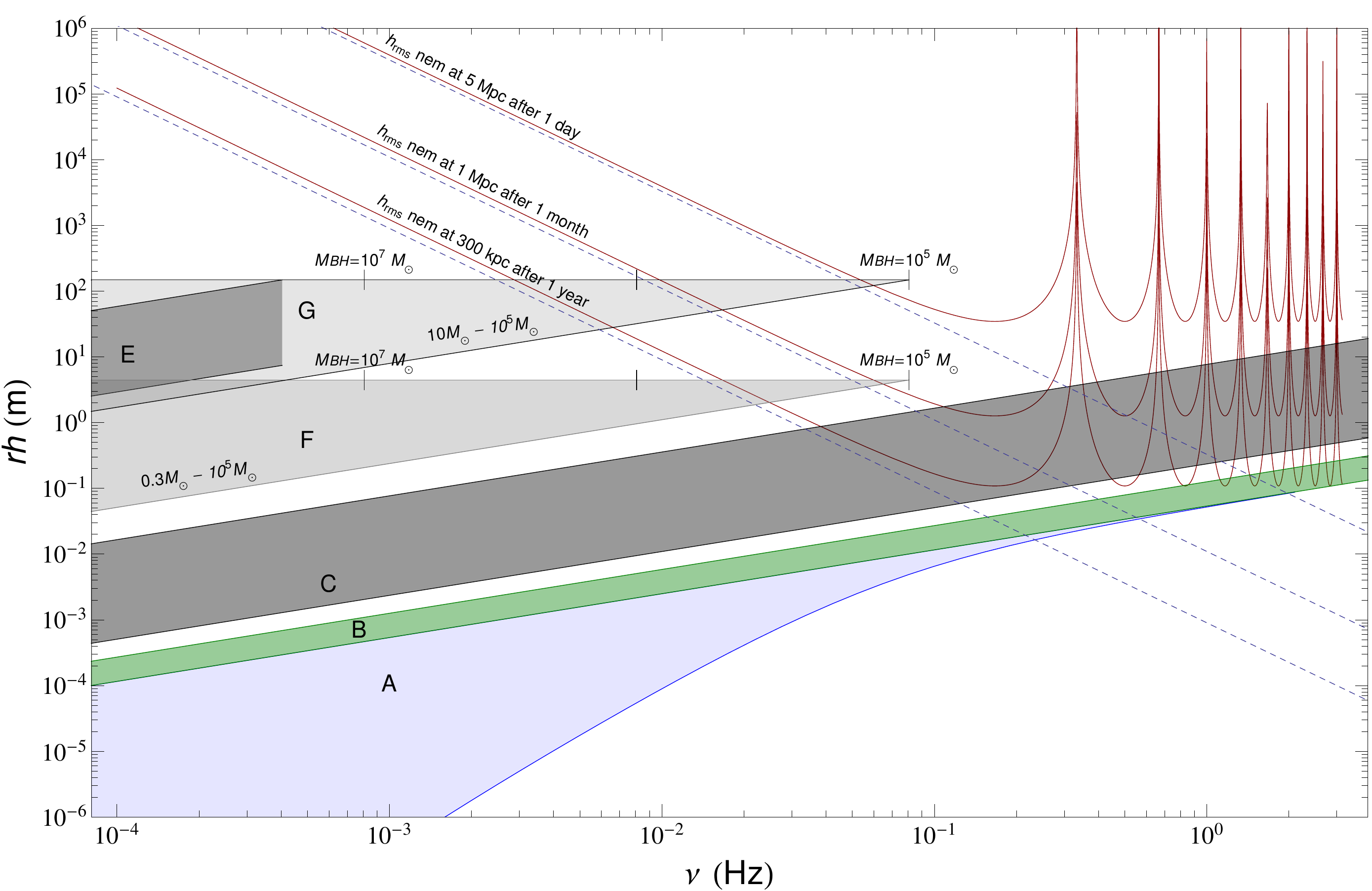}
   \caption{Signals from extragalactic sources. The triangular shaded regions F and G indicate the gravitational wave strain magnitude of
   extreme mass ratio inspirals (EMRIs) from binaries involving a supermassive black
   hole (SMBH) with ($M\geq 10^{5}\Msol$) and a smaller, compact partner.  The upper
   region G bordered in black corresponds to signals produced by a $10\Msol$ black hole
   in orbit about a supermassive partner, while the grey-bordered region F corresponds to the signal from a $0.3\Msol$ white dwarf in
such a system.  Vertical lines on the top horizontal border of the
EMRI regions indicate the point at which the binary orbit shrinks
to twice the gravitational radius of a SMBH with the indicated
mass.  The dark parallelogram E inset into the upper triangle
corresponds to the gravitational strain magnitude of EMRIs
involving the SMBH believed to lie at the center of Andromeda, for
smaller partners with mass between $0.5\Msol$ and $10\Msol$.  The
dashed blue lines indicate the point at which inspiraling bodies
double their frequency in less than one year (lower dashed line),
one month (middle) or one day (upper dashed line).  }
   \label{fig:EMCPlot}
\end{figure*}

\begin{figure*} %  figure placement: here, top, bottom, or page
   \centering
   \includegraphics[width=\textwidth]{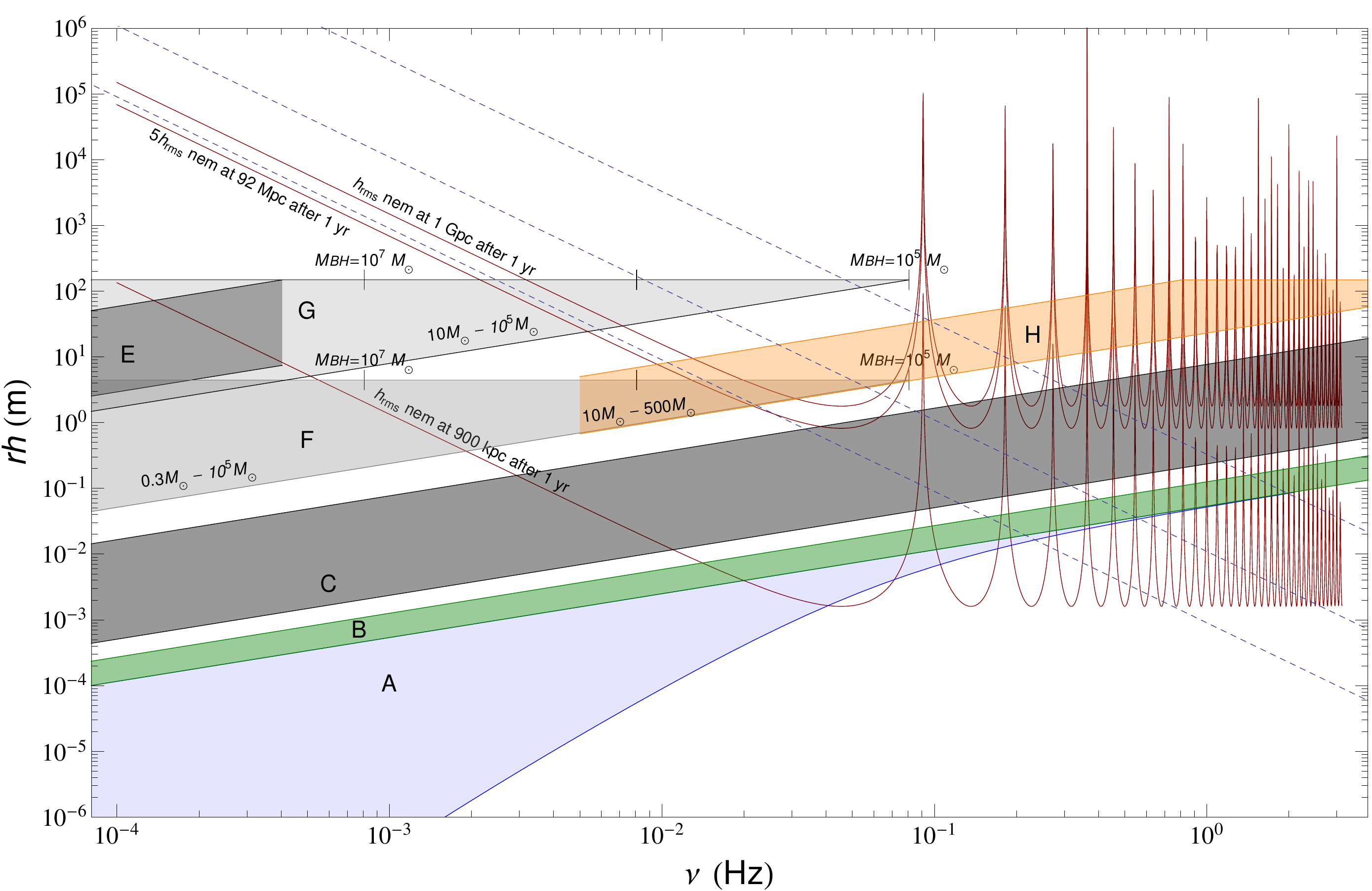}
   \caption{Signals from extragalactic sources seen with an optimized AGIS. All sources are as indicated in Fig.~\ref{fig:EMCPlot}.  An optimized AGIS would be sufficiently sensitive to observe inspirals at the core of Andromeda.  The dashed blue lines indicate the point at which inspiraling bodies double their frequency in less than one year (lower dashed line), one month (middle) or one day (upper dashed line).  The orange region H denotes the signal produced by inspirals of $10\Msol$ BH with a $500\Msol$ to $10^{4}\Msol$ partner within ESO 243-49.  Note that at distances approaching 1 Gpc, the effects of the cosmological redshift become significant, and this plot of source magnitude vs. frequency becomes inaccurate.  Regions A-G are as previously defined.}
   \label{fig:EMCPlotExtreme}
\end{figure*}

Intermediate mass ratio inspiral (IMRI) and extreme mass ratio
inspiral (EMRI) waves from other galaxies in the Local Group could lead to signal to noise ratios of larger than 1 albeit not above the minimum of 5 that we assume is required for detection, as indicated in Fig.~\ref{fig:EMCPlot}. Such sources might, however, be detectable at larger SNRs with the optimized AGIS.  Elements of the Local Group lie at distances as low as 22 kpc for
the nearer satellites of the Milky Way, out to distances in excess
of 1 Mpc.  There are two spiral galaxies other than the Milky Way
in the Local Group, Triangulum (M33) and Andromeda (M31), both of
which lie $\sim900$ kpc away~\cite{Ribas:2005}.  Using a detector with the
optimized parameters listed in Table~\ref{paramT} would extend
the range at which such systems could be detected to Gpc scales,
potentially allowing us to study the gravitational wave spectrum of the Local
Supercluster, as indicated in Fig.~\ref{fig:EMCPlotExtreme}. It would also be capable of detecting signals from
EMRIs involving the SMBH at the core of Andromeda, believed to
have mass $M=2\times 10^{7}\Msol$~\cite{Richstone:1990}.


%\subsection{Recent technological advances}
\section{Technology and Systematics\label{sec:techandsystem}}
We now turn to an analysis of the technologies required for the development of AGIS, and a discussion of various sources of systematic error.  Detection of gravitational waves will require hundreds or even thousands of
$\hbar k$ momentum splitting.  To date, large-momentum transfer has been accomplished via multiphoton Bragg diffraction, discussed in~\ref{multibragg}, to achieve smaller momentum splittings.  Although present interferometers are
very far from this level of performance, techniques involving accelerating optical lattices, such as the Bragg-Bloch-Bragg beam splitter, described in part \ref{BBBSec} and in more detail in~\cite{BBB}, may achieve the required
splittings.
AGIS will also require high atom throughput, the prospects for which are covered in part~\ref{sec:atomthroughput}.

Other technical requirements include the minimization of wavefront distortions in the laser beams. This will require a mode filtering cavity and high-quality optical elements
located inside the vacuum chamber.  Moreover, smooth frequency ramps for driving Bloch-oscillations are necessary, which can be generated by state of the art direct digital synthesizers. A fraction of the atoms may miss one or more of the momentum transfers and cause spurious interference, which must be eliminated in order to achieve shot noise limited signals.

\subsection{Atom Flux\label{sec:atomthroughput}}
AGIS will also require high atom throughput. Atomic fountains using
Raman sideband cooling have demonstrated launches of $2.5\times
10^8$ state selected atoms at a three-dimensional temperature of
150\,nK every two seconds by loading from a vapor cell MOT of
about $7\times 10^8$ atoms in a roughly 3\,mm-diameter cloud
\cite{Treutlein}. This flux can be increased by using a
two-dimensional MOT, which typically achieve a flux of about
$6\times 10^{10}$ atoms per second with a total laser power of
about 0.6\,W \cite{2DMOT}. Scaling linearly, we can expect
a flux of $10^{12}-10^{14}$ atoms with a laser power of
(10-1000)\,W, of which about 1/3 can be launched and cooled if the
efficiency of Ref. \cite{Treutlein} can be reproduced. Of course,
such scaling may not be straightforward: the laser power that can be achieved with a single commercial tapered amplifiers reaches about 2\,W, although 5\,W tapered amplifiers are under development \cite{PetersprivateComm}. The 2D-MOT beams can be generated by many such chips (as single mode beams are not required for 2D MOT operation), and the atom flux of several 2D MOTs can
be combined.

Increasing the atomic density beyond $\sim 3\times 10^9/$cm$^3$, demonstrated by Treutlein {\em et al.}, is undesirable because of mean field shifts \cite{meanfieldshift}, so a sample of $7\times 10^{12}$ will have a 10\,cm diameter. Such a diameter is compatible with the thick beams required for AGIS in order to achieve a large Rayleigh
range. Thus, while technical issues will need to be addressed, we
are confident that an atom flux of more than $\eta=10^{12}$
atoms/s can be achieved for AGIS.

Finally, atom detection methods with sufficient signal to noise ratios to detect $10^{12}$ atoms at the shot noise limit need to be developed. This may involve high solid angle light collection optics for fluorescence detection, the use of charge coupled devices (CCD) cameras and image processing techniques to suppress stray light.  Most importantly, laser frequency and intensity stabilization is required.  Chopping or modulating the beam at frequencies above the $1/f$ noise floor may be used to overcome technical noise.  Further sources of systematic error are discussed in more detail in part~\ref{sec:systematic}.

\subsection{Multiphoton Bragg diffraction\label{multibragg}}

Whereas classical atom interferometers use Raman transitions to
transfer the momentum of two photons to the matter waves,
multiphoton Bragg diffraction can be used to form large-momentum
transfer beam splitters. We have interfered cesium matter waves that were split by a momentum difference $\Delta p$ of up to
$24\hbar k$, the highest so far \cite{BraggInterferometry} (Other methods, also based on Bragg diffraction, have meanwhile achieved similar momentum splitting \cite{Domen}). The
sensitivity of interferometers rises proportional to $\Delta p$ in
measurements of inertial effects, e.g., local gravity, the gravity
gradient, or gravitational waves. The sensitivity even rises
proportional to $(\Delta p)^2$ in other applications, e.g., measurements of the fine structure constant $\alpha$ and the recoil frequency.

\subsection{Large momentum transfer by accelerated optical lattices\label{BBBSec}}

\begin{figure}
  \centering
  \includegraphics[width=3.5in]{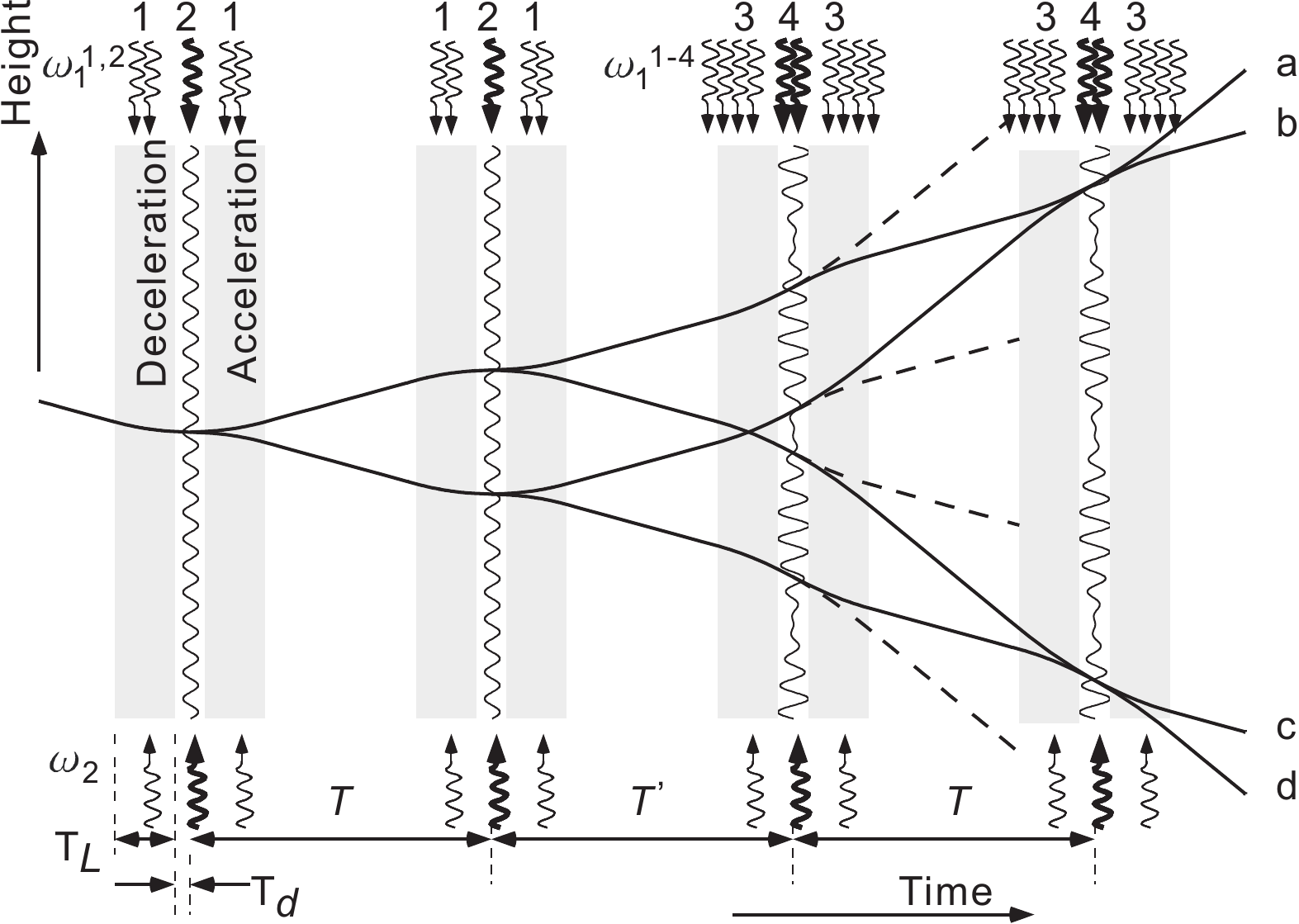}
  \vspace{-0.2cm}
  \caption{\footnotesize Space-time diagram of the
BBB-interferometers. 1: Bloch oscillations in dual accelerated
optical lattice; 2: Bragg beam splitter; 3: Bloch oscillations in
quadruple lattice; 4: Dual Bragg beam splitter.}
  \label{interferometer}
\end{figure}

With Bragg diffraction, increasing $\Delta p$ beyond
approximately $24\hbar k$ is difficult, because the
required laser power rises sharply with momentum transfer.
We overcome this limitation by coherent acceleration of
matter waves in optical lattices (Bloch oscillations, Ref.
\cite{Peik}) to add further momentum. With this Bloch-Bragg-Bloch
(BBB) beam splitter, we have achieved $\Delta p=88\hbar k$; in
simultaneous conjugate interferometers (SCIs) with $\Delta p=24\hbar k$, the BBB splitter allows us to see
interferences with $\sim 30\%$ of the theoretical contrast,
compared to $\sim 2\%$ with Bragg diffraction \cite{BBB}. It is worth mentioning that the SCIs with BBB splitters use, all in
all, 6 Bragg diffractions and 24 optical lattices, see Fig.
\ref{interferometer}. Since the BBB splitter does not require
higher laser power for increasing $\Delta p$, we expect that
technical improvements, discussed below, will allow us
to reach a splitting of hundreds or even thousands of photon
momenta. Gravitational wave detection aside, the BBB interferometer may also enable measurements of the Lense-Thirring effect \cite{Landragin}, tests of the
equivalence principle at sensitivities of up to $\delta g/g\sim
10^{-17}$ \cite{Dimopoulos}, atom neutrality \cite{Neutr}, or
measurements of fundamental constants with sensitivity to
supersymmetry \cite{Paris}.

Other necessary advances include high-power, ultra-low noise lasers
\cite{6Wlaser}, increased interferometer areas (and hence
sensitivity) \cite{Dimopoulos}, advanced algorithms for data
analysis \cite{ellipfit,Stockton}, and new atom-optics tools.

\subsection{Simultaneous Conjugate Interferometers}

The sensitivity of atom interferometers is often limited by the effects of vibrations. Using SCIs has
allowed us to cancel the influence of vibrations \cite{SCI}. This
allowed us to increase the time $T$ between light pulses to 50\,ms
from 1\,ms for atom interferometers with $\Delta p= 20\hbar k$,
which corresponds to an increase in sensitivity, by a factor of 50
for recoil frequency measurements and  2,500 for gravity
measurements.

\subsection{Systematic Influences\label{sec:systematic}}

So far, only atom shot noise has been considered in this paper.
Many other limiting influences have been studied \cite{GravWav},
and further studies will be required. This, however, is beyond the
scope of the present paper.

The following paragraphs mention some of the requirements for reaching the shot noise limit in the basic scenario of Tab. \ref{paramT}. This shot noise limit may be expressed as a 1\,$\mu$rad phase uncertainty per root Hz in the interference fringes of
the atomic matter waves, or, at a cycle time of $2T=6\,$s, about
0.4\,$\mu$rad in one cycle. For the optimized scenario, these figures
are 0.2$\,\mu$rad per root Hz and 0.04$\,\mu$rad per cycle,
respectively.

\subsubsection{Magnetic fields}

Atoms in $m_{F}$ = 0 quantum states only exhibit a quadratic
Zeeman effect, 427 $\mu$Hz/$(\textrm{mG})^{2}$ for Cs. If we
assume a background field of 1\,mG is applied to fix the
quantization axis, local fluctuations of 1\,$\mu$G=$10^{-11}\,$T
per root Hz will cause phase errors of less than $10^{-6}$rad per
root Hz, as required. This appears to be challenging. We are primarily concerned with fluctuations on the timescale of $T$, and much less so with a constant background. Magnetic field fluctuations on a time-scale of a second can be reduced to below $10^{-13}\,$T \cite{BudkerRomalis}. Thus in principle, magnetic fields can be kept
sufficiently low, even for the optimized scenario. It remains
challenging to achieve this over a tube length of
kilometers rather than centimeters.

\subsubsection{Gravity gradient noise}

Gravity gradient noise \cite{HughesThorne} is caused by moving objects, such as vegetation, the sea, the atmosphere, transportation, and most importantly, seismic activity. It cannot be shielded against, but it can be suppressed by choosing a location far away
from from major perturbations. Unfortunately, relatively little is
known about the exact magnitude of gravity gradient noise. To show how serious this problem is, we shall present a simple estimate of seismic noise relevant to the basic interferometer. We will assume that it is installed with its highest point 2 km below the surface.

A convenient starting point for this discussion is the ``new low noise model" for seismic noise on vertical component sensors presented in Fig. 1 of Ref. \cite{WidmerSchnidrig}. The dominant sources of noise are shown as being local atmosphere noise up to about 2\,mHz, hum due to high degree Earth normal modes near 3\,mHz, Rayleigh waves near 10\,mHz, and marine microseismics between 0.03 and 2\,Hz. According to Ref. \cite{Ekstrom}, Sect. 6, ``at quiet stations, much, and possibly most, of the vertical long-period seismic signal recorded in the period band 200 - 400\,s has the dispersive properties of fundamental-branch Rayleigh waves." For estimating the transfer function from vertical displacement noise in this frequency band, we will thus model all seismic activity as Rayleigh waves.

The vertical displacement due to a fundamental Rayleigh wave is given by \cite{Garland}
\begin{equation}
u = C\left(-0.85 e^{-qk z} +1.5 e^{-sk(x_3)}\cos k(ct-x)\right)
\end{equation}
Here, $z$ is the depth below the surface, $x$ is the horizontal direction of propagation, $k=2\pi/\lambda$ the wavenumber, and $c$ the speed of the wave. For an order of magnitude estimate, we use $c = 205$\,m/s, $q=0.88$ and $s=0.36$ as measured for the fundamental Rayleigh mode at the LIGO sites \cite{HughesThorne}. The second term is dominant because it is both larger and falls off more slowly with depth. Thus, we will assume the vertical displacement of matter is described by
\begin{equation}
u = b e^{-sk z}\cos k(ct-x),
\end{equation}
where $b$ is the surface vertical displacement.

\begin{comment}
Because the amount of matter is conserved, the equation of continuity demands
\be
\frac{\partial \rho}{\partial t}+\nabla\cdot(\rho v)=0
\ee
for the local density $\rho(x,z,t)=\rho_0+\delta \rho$ of soil, where $\rho_0$ is a constant and $v$ is the flow velocity. For harmonic time dependence, $\partial \delta \rho/(\partial t)=i \omega \delta \rho(x,y)$, so
\be
i\omega \delta \rho +\nabla\cdot(\rho_0 v)=0
\ee
We also use $v_z=\partial u/\partial t=-kc b e^{-skz}\sin k(ct-x)$:
\be
i\omega \delta \rho -\nabla\cdot(\rho_0 kc b e^{-skz}\sin k(ct-x)\hat z)=0
\ee

\be
i\omega \delta \rho +s \rho_0 k^2c b e^{-skz}\sin k(ct-x)=0
\ee

\be
\delta \rho =-\frac{s \rho_0 k^2c b e^{-skz}\sin k(ct-x)}{i\omega}=i s \rho_0 k b e^{-skz}\sin k(ct-x)
\ee

\end{comment}
From this, we can calculate the variation of the distance to the earth's surface: The acceleration of free fall at the surface of an infinite sheet of thickness $Z$ due to the sheet's mass is $a=\pi^2\rho G Z [1-(Z/R)/\pi+\ldots]$, where $R$ is the radius of the sheet. The effect of the seismic wave is to replace $Z\rightarrow z+u$, so $a=\pi^2\rho G (Z+b) [1-(Z/R)/\pi+\ldots]$. If the wavelength is infinite, this is going to have exact same effect on two atomic clouds and thus no effect on AGIS. To account for the finite wavelength, we use $\lambda \sim R$, and obtain a difference in accelerations of the top and bottom atomic clouds:
\begin{equation}
a_t-a_b=\pi \rho G b \frac{L}{\lambda}
\end{equation}
where the local density $\rho$ is about 1800\,kg/m$^3$ \cite{HughesThorne}. Finally, $\delta a$ has to be divided by $\omega^2$ and by the baseline length $L$ in order to obtain the resulting strain noise level in searching for GW signals,
\begin{equation}
h =  \frac{\rho G b}{2 \omega c}
\end{equation}
Using the new low noise model \cite{WidmerSchnidrig} we obtain, for example, $h=2\times 10^{-14}/\sqrt{\rm Hz}$ at $\omega=2\pi \times 10$\,mHz, and $h=1.5\times 10^{-16}/\sqrt{\rm Hz}$ at $\omega=2\pi \times 100$\,mHz. Both figures are between three and four orders of magnitude larger than the shot noise limit at 100\,mHz (920 times as large), and 10\,mHz (14,000 times as large), respectively, than the shot noise (Tab. \ref{paramT} and Fig. \ref{fig:strainnoise}). The problem is even more severe for the optimized detector.

This shows that gravity gradient noise is certainly one of the most important issues that need to be addressed to make AGIS viable. To suppress it, several strategies can be studies:
\begin{itemize}
\item Use of more than two atomic clouds, whose signals are correlated in order to suppress seismic noise while retaining sensitivity to gravitational waves.
\item Additional suppression of gravity gradient noise can be obtained by  seismic monitoring of the AGIS environment, but further study of the effects of seismically-induced gravitational gradients on an AGIS detector is needed. This can be expected to work best where it is needed most, for low frequency seismic waves. These waves have long correlation lengths and can be better modeled with a limited number of detectors.
\item Finally, setting up several AGIS can somewhat increase sensitivity through looking for correlations.
\end{itemize}
However, the above very crude and simplified estimates show that the problem needs the attention of an expert.

\subsubsection{Influence of laser noise}

While a detailed analysis has yet to be performed, there are two requirements on the laser noise. The first is on low frequency fluctuations on the time scale of the pulse separation time $T$, the second is for fluctuations on the time scale of the duration $t$ of the beam splitter pulse.

High-frequency laser noise must meet the following requirement: An interferometer with a pulse separation time $T$ at
the shot noise limit requires the phase uncertainly per beam
splitter be less than $\sim 1/(n\sqrt{2\eta T})$. If the beam
splitting pulse takes a time $\sigma$ (half width), it samples the
laser's phase noise spectral density over a bandwidth of roughly
$1/(2\pi \sigma)$. The resulting phase noise spectral density of the beam would be
\begin{equation}
\tilde \phi =\frac{\sqrt{2\pi\sigma}}{n\sqrt{2\eta T}},
\end{equation}
if the two interferometers were independent. However, since $\sigma \gg L/c$, this noise will mostly be common-mode to both interferometers. An estimate for the resulting requirement on the short term stability of the laser is thus
\begin{equation}
\tilde \phi =\frac{\sqrt{2\pi\sigma}}{n\sqrt{2\eta T}}\frac{\sigma}{L/c} \approx 10^{-5}/\sqrt{\rm Hz},
\end{equation}
where the basic parameters (Tab. \ref{paramT}) as well as $\sigma=100\,$ms were assumed. This is a mere -100\,dBc/$\sqrt{\rm Hz}$ phase noise which is easily achievable at the state of the art of lasers \cite{PLL}. Even the requirements for the optimized scenario ($1.6 \times 10^{-7}/\sqrt{\rm Hz}$, or -135\,dBc/$\sqrt{\rm Hz}$) can be met.

The low-frequency requirement arises since AGIS measures the distance between two atoms by comparing it to the phase of a standing wave of a laser. In order to reach the atom shot noise limit, fluctuations $\delta k$ in the effective wavevector must not fluctuate by more than the atom shot noise, $n (\delta k) L < 1/\sqrt{2 T \eta}$. The factor of $n$ exists because the effective wavenumber is, to leading order, $k_{\rm eff}=nk$. For the basic parameters in Tab. \ref{paramT}, this leads to $(\delta k)/k < 1/(nkL\sqrt{T \eta}) \approx 4\times 10^{-20}$. This corresponds to a $\sim 12\,\mu$Hz stability and must be maintained over the timescale of $2T$. No such lasers exist at present: the best cavity stabilizations \cite{BCY,Ludlow} reach Hz-level stability. See Ref. \cite{YemHz} for prospects for a mHz laser.

To illustrate this required level of stability, it is equivalent to the Doppler effect due to a $c/(nkL\sqrt{2T \eta}) \sim 1\times 10^{-11}$\,m/s velocity of the source. Thus, the position noise spectral density of the laser must not be more than
\begin{equation}
\tilde x=\frac{2Tc}{nkL\sqrt{\eta}}\approx 1.8\times 10^{-10}\,{\rm m}/\sqrt{\rm Hz}
\end{equation}
on the time scale of 2\,T. For comparison, the vibrational noise spectral density in the DUSEL underground facility was measured by Vuk Mandic (University of Minnesota) to be around $10^{-7}\,$m$/\sqrt{\rm Hz}$ around 0.1\,Hz, some three orders of magnitude short.

To compare that to the corresponding requirement in a light interferometer detector of the same sensitivity, we express $\tilde x$ as a function of the sensitivity $h$ in the low frequency limit: $\tilde x=ch_{\rm LF}T^3\omega^2$. By comparison, a light interferometer detector requires a mirror position noise spectral density of roughly $\tilde x_{\rm LIGO}=hL_{\rm LIGO}$. Thus, the requirement for AGIS is less stringent than the one of light interferometers by a factor of
\begin{equation}
\frac{\tilde x_{\rm AGIS}}{\tilde x_{\rm light}}=\frac{cT^3\omega^2}{L_{\rm light}}\sim 8\times10^7 \left( \frac{\omega}{2\pi\,{\rm Hz}}\right)^2
\end{equation}
for $L_{\rm light}=4\,$km and the basic scenario (the ratio is $4 \times 10^9[\omega/(2\pi\,{\rm Hz})]^2$ for the optimized scenario). Thus, while position noise requirements for AGIS are extremely stringent, they are much less stringent than those for light interferometers.

This points out a way in which the low frequency laser noise requirement can be alleviated: If it is possible to drive two AGIS sensors by the same laser, laser noise can be canceled. However, the beam splitting optics used must meet the vibration requirements just outlined. These requirements are much less stringent than corresponding requirements in LISA. Presenting a detailed scheme, however, is beyond the scope of the present work.

\subsection{Demonstrator Setup}

In order to demonstrate the critical technologies for GW
detection, we are assembling a tabletop demonstrator in our
lab. Our cesium setup is an atomic fountain, 1.5\,m tall~\cite{BraggInterferometry,SCI,BBB}, with a free
evolution time $T$ of 0.5\,s. Inside our atomic fountain tube,
we installed three layers of mu metal cylinders to suppress stray magnetic fields. The atoms are trapped in a
 two-dimensional magneto-optical trap (2D MOT), generated with
2\,W of laser power.  This system should be capable of producing a flux of $10^{11}$ atoms per second~\cite{2DMOT}.  Atoms from the 2D MOT are then
transferred to a 3D MOT, and subsequently launched by a
moving
optical molasses. The 3D MOT is about 2 cm in diameter and is estimated to contain on the order of $10^{10}$ atoms.  The 2D MOT and 3D MOT
chambers are separated by a differential pumping tube with a pressure ratio of $10^{3}$, in order to maintain the high loading rate of atomic flux without increasing the pressure in the main 3D MOT chamber, where we are able to maintain a pressure of about $10^{-10}$ Torr. The lifetime of the atomic sample is thus on the order of a few seconds, which is crucial for interferometry.  A temperature of less than $2\,\mu$K has been achieved on a daily basis by using polarization gradient
cooling and adiabatic cooling methods. Raman sideband cooling in an optical lattice~\cite{Treutlein} will
be used to further cool the atoms to
$\sim 350$\,nK in the $F=3,
m_F=-3$ quantum state. At this temperature, the atomic sample will
 expand to about 1\,cm over the course of the experiment.
The atoms
will be transferred to the $m_F=0$ state by applying a small
magnetic bias field and a 10\,W microwave sweep. A velocity
selective Raman
transition will reduce the vertical velocity width
to 0.3 recoil velocities. The BBB beam splitter will be optimized for detection of gravitational waves.
The aim is to achieve the high momentum splitting required for AGIS.

Our laser system for driving Bloch oscillations and Bragg
diffraction will be based
on a 6\,W Ti:sapphire laser
\cite{6Wlaser,SCI} (that we constructed from a Coherent 899
laser without intracavity etalons) and a highly efficient
arrangement
of acousto-optical modulators.

\section{Summary and Outlook}

We have studied candidate sources of gravitational waves for detection by an atomic gravitational wave interferometric sensor, AGIS. We describe an optimized set of experimental parameters for reaching high sensitivity in a free-falling implementation of AGIS, where the atoms do not interact with the laser beams except during beam splitter operations.

We consider binary inspirals to be the best candidates for detection by AGIS.  A ground-based AGIS may be capable of detecting type Ia supernova precursors at up to 500\,pc.  An optimized detector should observe gravitational waves from at least one presently known binary system.  AGIS may detect all neutron star-neutron star binary coalescences within 16 kpc, some up to a year in advance.  It would also be sensitive to neutron star-black hole mergers within 50 kpc.  AGIS may be able to observe the final decades of inspiraling black holes at the galactic core, and potentially the final years of an in-falling neutron star.  AGIS would also be able to observe the final days of some extreme mass ratio inspirals within the Local Group, although it is uncertain that there are any supermassive black holes in the appropriate mass range. However, current estimates for the rate of occurrence of several classes of detectable events are spread over a disquietingly large range.

Based on the shot noise limit, an optimized AGIS may be able to observe the late stage inspirals at the supermassive black hole in Andromeda, and to more general extreme mass ratio inspirals within the Local Supercluster, although we caution the reader that many other noise sources, such as Newtonian noise, may preclude this and remain to be studied.  AGIS would then be useful for probing gravitational wave emissions from sources at non-negligible redshifts.

However, several technical challenges need to be met in order to reach this: 
\begin{enumerate}
 \item ultrahigh coherent momentum transfer to the atoms,
 \item high atom flux of $10^{12}/$s,
 \item overcoming wavefront distortions in the laser beams.
 \item high-power lasers will need to be developed.
 \item gravity gradient noise
 \item laser phase noise.
 \end{enumerate}
Schemes for overcoming these disturbances need to be developed. These include a multi-arm version of AGIS, in which the effects of laser noise cancel, the use of mHz-linewidth lasers. Seismic noise seems to be about 3 orders of magnitude larger than the shot noise of even basic AGIS, though strategies for reducing this might exist. For example, seismic monitoring and underground operation is currently being studied as a means of overcoming gravity gradient noise. This issue will definitely require the attention of experts. In addition, it has been pointed out \cite{Bender:2010} that wavefront distortions in the laser beams together with shot-to shot fluctuations in the location of the atomic cloud give rise to additional noise. 

Finally, we note that even our optimized scenario could be surpassed by more advanced atom-optics methods. For example, by holding the atoms in an optical lattice throughout the interferometer cycle, the pulse separation time can be extended beyond 11\,s.

\acknowledgements We thank Phillipe Bouyer, Mark Kasevich and Vuk Mandic for several important discussions at the 2009 DUSEL workshop in Lead, SD.  We also thank Peter Bender, Flavio Vetrano and Guglielmo Tino for useful suggestions.  This work is supported, in part, by a precision measurement grant of the National Institute of Standards and Technology, the Alfred P. Sloan Foundation, and by the David and Lucile Packard Foundation.


\begin{thebibliography}{99}
\bibitem{meanfieldshift} The mean field shift at this density is about 10\,mHz  for the Cs clock transition \cite{Gibble}; its influence on atom interferometry can theoretically be eliminated by using same internal states, but in practice fluctuations in the beam splitter efficiency will preclude perfect elimination.
\end{thebibliography}
\end{document}